\documentstyle[a4,12pt,epsfig,graphicx]{article}
\begin{document}
\titlepage
\begin{flushright}
CERN-TH/2001-137\\
DAMTP-2001-36\\
T/01-056\\
hep-th/0105269 \\
\end{flushright}
\vskip 1cm
\begin{center}
{ \Large
\bf On Brane Cosmology and  Naked Singularities}
\end{center}

\vskip 1cm
\begin{center}
{\large Ph. Brax\footnote{email: philippe.brax@cern.ch} }
\end{center}
\vskip 0.5cm
\begin{center}
Theoretical Physics Division, CERN\\
CH-1211 Geneva 23\footnote { On leave of absence from  Service de Physique Th\'eorique, 
CEA-Saclay F-91191 Gif/Yvette Cedex, France}\\
\end{center}
\vskip .2 cm
\begin{center}
{\large A. C. Davis\footnote{email: a.c.davis@damtp.cam.ac.uk}}
\end{center} 

\vskip .5 cm
\begin{center}
DAMTP, Centre for Mathematical Sciences, Cambridge University,
Wilberforce Road, Cambridge, CB3 0WA, UK. 
\end{center}

\vskip 2cm
\begin{center}
{\bf  Abstract}
\end{center}
\vskip .2 cm
\noindent

\vskip .3in \baselineskip10pt{
Brane-world  singularities are analysed,  emphasizing the case of 
supergravity in singular spaces where the singularity puzzle is naturally
resolved.
These naked singularities are either time-like or null, corresponding to the finite or infinite
amount of conformal time that massless particles take in order to reach them. 
Quantum mechanically we show that the brane-world 
naked singularities are inconsistent. Indeed we find 
that time-like singularities are not wave-regular, so 
the time-evolution of wave packets is not uniquely defined in their vicinity,
while null singularities  absorb incoming radiation. 
Finally we stress that for supergravity in singular spaces 
there is a topological 
obstruction, whereby naked singularities are necessarily screened off by the 
second boundary brane. 
}
\bigskip
\vskip 1 cm

\noindent
\newpage
\baselineskip=1.5\baselineskip
\section{Introduction}
Brane cosmology\cite{bine} has recently appeared as a framework where thorny issues such as
the cosmological constant problem can be tackled. In particular new mechanisms
have been proposed whereby the vacuum energy of the brane-world curves
the fifth dimension leaving a flat four dimensional brane-world intact\cite{dimopoulos}.
Unfortunately this scenario seems to fail as the existence of a naked singularity 
in the bulk prevents one from obtaining a smooth five dimensional space-time.
This singularity needs to be resolved leading to a fine-tuning between the
tension on the brane-world and of the ghost brane sitting at the singularity\cite{sing}.

Another proposal involves supergravity in singular spaces\cite{kallosh}. In that case $N=2$
supergravity lives in the bulk and is broken on the brane-world. The breaking
of supersymmetry ensures that a non-static configuration is generated\cite{us}. The resulting
brane-world metric is of the FRW type with an acceleration parameter $q_0=-4/7$
and an equation of state $\omega=-5/7$, within the experimental ball-park\cite{supernova}. 
Unfortunately a thorough analysis of the coupling of supergravity in singular spaces
to ordinary matter on the brane-world shows that the amount of supersymmetry breaking
needs to be fine-tuned to the level of the critical energy density of the universe\cite{cosm}.    
Nevertheless this model is relevant as
it realizes in an explicit way a quintessence scenario \cite{quintessence}in five dimensions. 
As in the self-tuned brane scenario
there is a would-be singularity in the bulk. One of the aims of this paper is
to provide a description of how a topological obstruction prevents
the existence of such a naked singularity in the bulk. 

In a first section we shall study the classical trajectories
of massive and massless particles by studying the geodesics in
warped geometries with naked singularities. For massless particles the
singularity can be reached in either a finite  or an infinite  amount of conformal
time corresponding to time-like or null singularities. The former corresponds to the self-tuned brane while the latter
appears for supergravity in singular spaces.
We then study the quantum mechanics of gravitons and show that the time-like naked 
singularities are {\it repulsons}\cite{linde},
repelling all incoming radiation. We also find that they are
not wave-regular\cite{hor,ishi,brax}, i.e. the time evolution of wave packets is not uniquely
defined, prompting the necessity of imposing an appropriate boundary condition
at the singularity, i.e. knowing enough about its possible resolutions. On the contrary,
null singularities are wave -regular but absorb incoming radiation.
In both case this signals a quantum inconsistency of the models. 
Fortunately,  in the supergravity case, the would-be singularity is absent
due to a topological obstruction associated with the presence of 
four-forms in the bulk. This obstruction is reminiscent of the tadpole
cancellation mechanism in string theory\cite{pol,luk}.

\section{Geodesic Motion}

In this section we are interested in the classical  motion of massless and massive particles
in the vicinity of the  naked singularities arising in brane-world scenarios.
More precisely we consider the bulk geometry to be described by the metric
\begin{equation}
ds^2= a^2(u)(du^2 -d\eta^2 + dx_idx^i)
\end{equation}
in conformal coordinates.
The behaviour of the scale factor $a(u)$ is given by a power law close to  
the singularity
\begin{equation}
a(u)=(\frac{u}{u_0})^{\beta}. 
\end{equation}
The curvature vanishes at the origin for $\beta>-1$ and at infinity for $\beta<-1$.
We will investigate both situations in the following.

The classical motion is characterized by the point-particle Lagrangian\cite{gro}
\begin{equation}
{\cal L}\equiv \frac{ds^2}{d\tau^2}=a^2(u)(\dot u^2 -\dot \eta^2 +\dot x^i \dot x_i).
\end{equation}
The time and space independence of the Lagrangian leads to the conservation of energy and momentum
\begin{equation}
E=-2a^2(u)\dot \eta,\ \ p^i=2a^2(u) \dot x^i
\end{equation}
thus giving the reduced Lagrangian
\begin{equation}
{\cal L}=a^2(u) (\dot u^2 + \frac{p^2-E^2}{4a^4}).
\end{equation}
The  trajectories are determined by the constraint
\begin{equation}
{\cal L}=\epsilon
\end{equation}
where $\epsilon=0$
for light-like paths and  $\epsilon=-1$ for time-like paths.
Let us consider a particle sitting initially on the  brane-world with speed
$\dot u\vert_{0}$.
We can rewrite the Lagrangian constraint as
\begin{equation}
\dot u^2 +\frac{\epsilon-\dot u\vert_{0}^2}{a^4} -\frac{\epsilon}{a^2}=0.
\end{equation}
This is the classical motion of a massive particle with zero total energy in the
potential 
\begin{equation}
V(u)=\frac{\epsilon-\dot u\vert_0^2 }{ a^4(u)} -\frac{\epsilon}{a^2(u)}.
\end{equation}
First of all notice that the particles with a vanishing initial velocity
$\dot u_0=0$ remain on the brane. 

Let us now consider the geodesics obtained by launching the particles
from the brane to the singularity.
For $\beta>0$ the potential
goes to minus infinity at the singularity and vanishes at infinity. For $-1<\beta <0$ the
potential vanishes at the singularity and goes to minus infinity at infinity.
For $\beta<-1$ the potential goes to minus infinity at the singularity.
There is a critical point at 
\begin{equation}
 a^2_*=2+ 4\dot u\vert_{0}^2
\end{equation}
in the massive case.
For $\beta>0$ the critical point is beyond the brane-world located at 
$u_0$ while
for $\beta<0$ it is between the brane-world and  the origin.

Let us consider massive particles first. In the case $\beta>0$, 
as the total energy vanishes and the potential energy of the
particle is always negative,  we find that  massive particles  are  
attracted by the singularity.
In the case $-1<\beta<0$ massive particles  evolve  
in the bulk before reaching a  turning point for 
\begin{equation}
a_{tp}^2=1+\dot u_0^2.
\end{equation}
Notice that this esures that massive particles never reach the singularity 
in the $-1<\beta <0$ case. For $\beta <-1$ the massive particles are attracted by 
the singularity.

Massless particles are attracted by the singularity. Indeed the geodesics are given by
\begin{equation}
u=u_0 -\frac{2\dot u_0}{E} \eta
\end{equation}
showing that the singularity is reached in a finite 
amount of conformal time
for $\beta>-1$,  corresponding to a time-like  singularity
whereas for 
$\beta < -1$ the amount of conformal  time is infinite, i.e. a null singularity. 

Two particularly relevant cases have been discussed lately. First of all
the self-tuned brane scenario\cite{dimopoulos} is such that $\beta=1/3$.
This corresponds to an attractive time-like singularity. All kinds of matter, whether massive
or massless, are attracted and reach the singularity in a finite amount of conformal time.
As such this singularity does not make sense classically. It has been argued
that it needs to be resolved by putting an appropriate brane located at the singularity
whose tension compensates for the tension of the original brane at $u_0$.

Another scenario invokes the presence of supergravity in the bulk with broken
supersymmetry on the brane\cite{us}. There is a would-be singularity whose existence
will be further discussed in section 4. It is characterized by $\beta =-3/2$.
This corresponds to a null  singularity.

In the next section we will study the quantum mechanical behaviour of massless
particles in the vicinity of such   naked singularities.

\section{Wave-Regularity of Naked Singularities}

We have seen that the brane-world singularities attract massless particles. It is then relevant to analyse their quantum mechanical behaviour 
in the neighbourhood of the naked singularity. We shall restrict our attention
to massless particles in the s-wave channel.
We assume that the only massless particle propagating in the bulk is the graviton.
We are interested in gravitons polarized along the brane-world.
The graviton wave function can be written as
\begin{equation}
h_{ij}= H(u,x) \epsilon_{ij}
\end{equation}
where $H(u,x)=H(u) e^{ik.x}$, $\epsilon_{ij}$ is the polarization tensor and $k$ is in the time direction.
In the Einstein frame the graviton equation reduces to the Laplace equation
\begin{equation}
\Delta h_{ij}=0.
\end{equation}
The polarization tensor must be traceless $\eta^{ij}\epsilon_{ij}=0$
and transverse to $k$ implying that $\epsilon_{0i}=0$. Denoting
by $\tilde\epsilon$ the spatial part of the polarization tensor we find
a basis of these tensors given by off-diagonal symmetric matrices
with $\tilde \epsilon_{ab}=\tilde \epsilon_{ba}=1$ and zero otherwise
along with diagonal matrices such that $ \tilde \epsilon_{aa}=1,\
\tilde \epsilon_{bb}=-1,\ a<b$. The latter are diagonal
polarizations while the former are transverse polarizations.
In the diagonal case put
\begin{equation}
H(u,x) =a \phi(u,x) .
\end{equation}
Then the scalar field $\phi$ satisfies the free wave equation
\begin{equation}
\nabla_{\mu}\nabla^{\mu} \phi=0.
\end{equation}
In the transverse case the function $H(u,x)$ satisfies the free scalar equation too.

In the following we shall concentrate on the scalar wave equation in five dimensions.
It is particularly useful to introduce
\begin{equation}
\phi=a^{-3/2}\psi
\end{equation}
which satisfies the  Schrodinger equation 
\begin{equation}
\psi''-V\psi =0
\end{equation}
where
\begin{equation}
V=-\omega^2 +\frac{(a^{3/2})''}{a^{3/2}}.
\end{equation}
It is possible  to recast the Schrodinger equation into the form\cite{brax} 
\begin{equation}
(\bar Q Q-\omega^2) \psi=0
\end{equation}  
where
\begin{equation}
Q=-\frac{d}{du}+\frac{1}{2}\frac{d\ln a^{3}}{du},\ \bar Q=\frac{d}{du}+
\frac{1}{2}\frac{d\ln a^{3}}{du}.
\end{equation}
The Hamiltonian $\bar Q Q$ is a symmetric operator $(f,Q\bar Qg )=(Q\bar Q f,g)$,  where
\begin{equation}
(f,g)=\int du f^*(u) g(u)+ \int du D_{u}f^*(u)D_{u} g(u),
\end{equation}
for functions depending only on $u$
if one restricts the domain  of $\bar Q Q $ to the infinitely differentiable
functions with compact support. 
Following \cite{ishi}we choose a Sobolev norm as it is related
to the energy of the scalar field $\phi$. In particular fields with finite norm have  finite energy.
With this choice the Hamiltonian is symmetric but is not guaranteed to be a self-adjoint operator\cite{hor,ishi}.
Notice too that the Hamiltonian
is a positive operator with two zero modes
\begin{equation}
\psi_1(u)=a^{3/2},\ \psi_2(u)= a^{3/2}\int ^{u}\frac{dv}{a^3(u)}.
\end{equation}
The rest of the spectrum is positive preventing the existence of tachyons.

We can solve the Schrodinger equation corresponding to the brane-world singularities as
\begin{equation}
V=-\omega^2 +\frac{3\beta}{2}(\frac{3\beta}{2}-1)\frac{1}{u^2}.
\end{equation}
The self-tuned brane scenario with $\beta=1/3$ and the supergravity scenario with $\beta=-3/2$ lead to  attractive 
singularities as the potential decreases at the origin in the former case, 
and at infinity in the latter case respectively. 
It is convenient to define $z=\omega u$. The generalized eigenstates read
\begin{equation}
\psi^1_{\omega}(z)=\sqrt{z} J_{(3\beta -1)/2}(z), \ \psi^2_{\omega}(z)=\sqrt z J_{(1-3\beta)/2}(z)
\end{equation}
for $\beta\ne 1/3$. In the latter case, i.e. for the self-tuned brane scenario,
the solutions  are  expressed in terms 
of zeroth order Bessel and Neumann
functions
\begin{equation}
\psi^1_{\omega}(z)=\sqrt{z} J_{0}(z), \  \psi^2_{\omega}(z)= \sqrt zN_{0}(z).
\end{equation}
Close to the time-like singularity the constant term in the potential is negligible implying that all the solutions
behave like the two zero modes
\begin{equation}
\psi_1(z)=z^{3\beta/2}, \ \psi_2= z^{1-3\beta/2}
\end{equation}
for $\beta\ne 1/3$ and
\begin{equation}
\psi_1(z)=\sqrt z,\ \psi_2(z)= \sqrt z \ln z.
\end{equation}
for $\beta=1/3$.
As none of the eigenstates are  oscillatory in the neighbourhood of the time-like singularity this implies 
that  
no flux reaches it. This is natural when the singularity is repulsive. For attractive singularities this is due to the
extreme steepness of the potential. Such singularities are {\it repulsons}\cite{linde}. 
For $\beta<-1$ the singularity is at infinity where the eigenfunctions behave like plane waves.
This implies that in a scattering experiment there will be some absorption by the
null singularity.  

We can now study  whether the time evolution of wave packets is well defined
in the vicinity of the naked singularities. 
To do that let us write the massless Klein-Gordon equation 
in the form
\begin{equation}
\frac{\partial^2 \phi}{\partial t^2}=- M \phi
\end{equation}
where $M$ is a second order partial differential operator depending only on the spatial derivatives.
After a change of variable, $M $ reduces to the Hamiltonian $\bar Q Q$. 
The Klein-Gordon equation defines a unique time evolution provided it can be written in the Schrodinger form
\begin{equation}
\frac{\partial \phi}{\partial t}=i M^{1/2} \phi
\end{equation}
for a unique self-adjoint operator $M^{1/2}$. 
This is equivalent to finding a unique self-adjoint extension to the symmetric operator $\bar Q Q$, i.e.
the Hamiltonian $\bar Q Q$ is essentially self-adjoint.

For null singularities at infinity, there is a single self-adjoint extension
of the symmetric operator $\bar Q Q$  acting on functions decreasing fast enough at infinity\cite{hor}. Hence the time-evolution of wave packets is well-defined. 
For the time-like case 
there is a useful criterion
of essential self-adjointness\cite{hor,ishi}.
Let us consider the eigenvalue problem
\begin{equation}
\bar Q Q \psi =\pm i\psi .
\label{index}
\end{equation}
It reduces to a Schrodinger problem
in a complex potential
\begin{equation}
\tilde V=V \pm i .
\end{equation}
Denote by $n_{\pm}$ the number of normalizable solutions  to (\ref{index}).
As the operator $\bar Q Q $ is real one has $n_+= n_-$, implying that there 
always exists 
self-adjoint extensions. Now the operator is essentially self-adjoint provided
$n_{\pm}=0$, i.e. the solutions are not normalizable.
Due to the finiteness of the fifth dimension, the only possible source of divergence
is at the singularity. Therefore one must check whether or not the 
solutions of (\ref{index}) are normalizable close to the singularity. 

In our case notice that in the vicinity of the singularity the extra complex term to $\tilde V$ is  negligible,
implying that the solutions are expressed in terms of the two zero modes\footnote{
The case $\beta=2/3$ is special as the potential vanishes. It is easy to see
that the eigenfunctions are normalizable.}. 
The issue of the quantum mechanical behaviour  of the singularity is
now dependent on the norm of these  eigenfunctions. 
Using the fact that the covariant derivative of the metric vanishes, 
we find that the norm of $\psi_1$ is finite provided
\begin{equation}
\int du\  a^3 < \infty
\end{equation}
which leads to 
\begin{equation}
\beta >  -\frac{1}{3}.
\end{equation}
Similarly the norm of $\psi_2$ is finite provided 
\begin{equation}
\int du \frac{1}{a^3} < \infty
\end{equation}
leading to
\begin{equation}
\beta < \frac{1}{3}.
\end{equation}
Therefore we find that there is always one of the zero modes 
which is normalizable.
This implies that the Hamiltonian is not essentially self-adjoint.
Hence we cannot define a unique evolution operator in the neighbourhood of the singularity.
Uniqueness of the evolution operator can be achieved if a physical
choice of boundary condition at the singularity is imposed. This requires 
more knowledge about the physics of the singularity, i.e. its resolution.

We have thus shown that the quantum mechanical behaviour of brane-world singularities
is pathological. Indeed the time-like singularities are repulsons while not wave-regular.
This requires knowledge about the resolution of the singularity in order to define
the time evolution of wave packets in their vicinity. On the contrary null singularities
are wave regular allowing one to study the evolution of wave packets in their vicinity
irrespective of the physical nature of the singularity. Unfortunately the null
singularities have a non-vanishing absorption cross section which needs to be
interpreted in order to make sense. In particular this absorption
might signal the presence of fields at the null singularity to which the
gravitons couple. In any case this requires a deep undertanding of the nature of the
singularity .
The time-like case is exemplified by the self-tuned brane models while the null case
occurs for supergravity in singular spaces. 
In the next section we will study the resolution of naked singularities in singular space
supergravity.

\section{Supergravity in Singular Spaces}

We have seen that the brane-world naked singularity are either not wave-regular, so that the time
evolution operator is not well-defined in their neighbourhood, or they absorb 
incoming radiation.
This is an inconsistency of the models which needs to be cured. In the following we shall treat the
case of supergravity in singular spaces \cite{kallosh} where such singularities may  occur.
Nevertheless we will show that there is a topological obstruction to the presence of naked singularities in the bulk.

Let us first discuss the on-shell
bosonic part of the Lagrangian
\begin{equation}
S_{bulk}=\frac{1}{2\kappa_5^2}\int \sqrt {-g_5}(R-\frac{3}{4}(g_{ij}\partial_{\mu}
\phi^i\partial^{\mu}\phi^j+V))
\label{lag}
\end{equation}
for a particular sigma model metric $g_{ij}$.
The bulk potential is given by
\begin{equation}
V=W_iW^i-W^2.
\end{equation}
as a function of the superpotential $W$.
On shell the bosonic Lagrangian   (\ref{lag}) is supplemented with  the 
boundary term  
\begin{equation}
S_{bound}=-\frac{3}{2\kappa_5^2}\int d^5x (\delta_{x_5}-\delta_{x_5-R})(\sqrt {-g_4}W).
\end{equation}
In the case of a single scalar field in the bulk the superpotential reads 
\begin{equation}
W=\xi e^{\alpha \phi}
\end{equation}
where $\xi$ is a scale.
The corresponding solutions in conformal coordinates are given  by\cite{us}
\begin{equation}
a=(\frac{u}{u_0})^{1/(4\alpha^2-1)}.
\end{equation}
Supergravity implies that $\alpha=1/\sqrt 3, \ -1/\sqrt {12}$. 
For the latter we find that $\beta=-3/2$.

Fortunately the bulk singularity is forbidden by the off-shell formulation of
supergravity in singular spaces.
The off-shell theory 
depends on two new fields. There is a supersymmetry singlet  $G$
and a four form $A_{\mu\nu\rho\sigma}$. One also introduces a modification of the bulk action by
replacing $g\to G$ and adding  a direct coupling
\begin{equation}
S_A=\frac{1}{4! \kappa_5^2}\int d^5x \epsilon^{\mu\nu\rho\sigma\tau}A_{\mu\nu\rho\sigma}\partial_{\tau}G.
\end{equation}
The boundary action is taken as
\begin{equation}
S_{bound}=-\frac{1}{\kappa_5^2}\int d^5x (\delta_{x_5}-\delta_{x_5-R})(\sqrt {-g_4}\frac{3}{2}W +\frac{2g}{4!}\epsilon^{\mu\nu\rho\sigma}A_{\mu\nu\rho\sigma}).
\label{tune}
\end{equation}
The supersymmetry singlet $G$ satisfies a first order constraint
\begin{equation}
\partial_{x_5} G=2g (\delta_{x_5}-\delta_{x_5-R})
\label{top}
\end{equation}
where the left hand side is nothing but the $A_{\mu\nu\rho\sigma}$  charge 
associated with the boundary branes. 
The constraint (\ref{top}) leads to the topological obstruction of singularities in the bulk.
Indeed from
\begin{equation}
\int dx_5 \partial_{x_5} G=0
\end{equation}
due to the compactness of the fifth dimension we deduce that the total charge
in the extra dimension much vanish. This is the equivalent to Gauss' law, or 
the tadpole cancellation mechanism in M-theory and string theory.
If one were to have a singularity in the bulk, the total charge would not 
vanish unless the singularity carries a charge, i.e. the extreme case where 
the singularity sits at the second brane. 
In all other cases the topological obstruction requires that the singularity 
be screened off by the second brane.

The same mechanism is at play when supersymmetry is broken on the brane
by detuning one of the tensions. In that case the boundary Lagrangian becomes
\begin{equation}
S_{bound}=-\frac{1}{\kappa_5^2}\int d^5x \delta_{x_5}(\sqrt {-g_4}\frac{3T}{2}W +\frac{2g}{4!}\epsilon^{\mu\nu\rho\sigma}A_{\mu\nu\rho\sigma})
\end{equation}
on the non-supersymmetric brane. The supersymmetry breaking parameter is $T\ne 1$. The Lagrangian on the supersymmetric brane is not modified. As the brane 
charge is not modified, the first order constraint (\ref{top}) remains leading,
to the same topological obstruction as in the supersymmetric case.
So supergravity in singular spaces leads to a natural resolution
of the singularity puzzle.

However it may appear that the tension on the
second brane  (\ref{tune}) has been fine-tuned to the opposite value
of the tension of the brane-world. As such this would be 
a phenomenon akin to the
one presented in  \cite{sing} where the ghost brane and the brane-world
have the same tension. In the case of supergravity in singular spaces 
the mechanism is more subtle. Indeed Gauss's law implies that the total charge
vanishes. Hence , by supersymmetry, this leads
to the vanishing of  the total tension. Now when supersymmetry is broken on 
the brane-world  by modifying  the tension of the brane-world, 
we lose the exact cancellation
between the brane tensions. Nevertheless the total charge still has to vanish,
implying that the would-be singularity is screened off by the second brane.

\section{Conclusions}

In this paper we have analysed the naked singularities inherent in 
self-tuned branes or the supergravity in singular spaces. These theories
have been put forward as possible five-dimensional explanations to the 
cosmological constant problem. We have shown that, for supergravity in
singular spaces, the singularity problem resolves itself. This is due to
the existence of a topological obstruction, requiring by Gauss's law that the 
total charge vanishes. Thus in this theory the singularity {\it must} lie
beyond the second brane, unless the singularity itself carries a charge,
in which case it sits at the second brane. Thus there is a natural resolution
to the singularity puzzle in this theory. 

For the self-tuned brane there is no such natural resolution. We have
analysed the behaviour of the singularity both classically and quantum
mechanically. Classically the singularity attracts massless particles,
in  a finite amount of conformal time, this being due
to the behaviour of the scale factor close to the singularity.
 Quantum mechanically we have shown that the  singularity is a
{\it repulson}, reflecting all incoming radiation. However it is 
not wave-regular, so that time evolution of wave packets
is not uniquely defined in the vicinity of the singularity. 

Our results suggest that the theory of supergravity in singular spaces 
is a well defined theory cosmologically. We have previously shown that
this theory leads to a natural cosmological evolution of the universe,
with a late stage of acceleration and cosmological constant consistent
with experiment\cite{supernova}. Thus this model deserves further investigation\cite{us&carsten}.

\section{Acknowledgements}

This work was supported in part by PPARC. PhB thanks DAMTP and ACD thanks
CERN for hospitality while this was in progress. 

\def\Journal#1#2#3#4{{#1}{\bf #2}, #3 (#4)}
\def\NPB{Nucl.\ Phys.\ {\bf B}}
\def\PLB{Phys.\ Lett.\ {\bf B}}
\def\PRL{Phys.\ Rev.\ Lett.\ }
\def\PRD{Phys.\ Rev.\ D }
\def\JPA{J.\ Phys.\ {\bf A}}
\def\JHEP{JHEP}

\end{document}